\documentclass[usenames,dvipsnames,screen,format=sigconf,table,xcdraw,10pt]{acmart}
\renewcommand\footnotetextcopyrightpermission[1]{} 
\usepackage{enumitem}
\setitemize{leftmargin=4mm}
\setenumerate{leftmargin=4mm}
\setlist{noitemsep,topsep=0pt,parsep=0pt,partopsep=0pt}

\usepackage[subtle, lists=tight]{savetrees}

\usepackage{caption}
\usepackage{subcaption}
\usepackage{pgfplots}
\usepgfplotslibrary{external}
\pgfplotsset{compat=1.18}

\newcommand{\eat}[1]{{}}

\usepackage[justification=RaggedRight,labelfont=bf,font=small,textfont=normalfont,tableposition=below]{caption}
\setitemize{leftmargin=4mm}

\begin{document}

\title{\huge Robustifying Measurement-Based Congestion Control Algorithms}
\author{Zhu Yuxi, Meng Zili, Shen Yixin, Xu Mingwei, Wu Jianping}

\renewcommand{\shortauthors}{Zhu Yuxi}

\begin{abstract}

The design methodology of congestion control algorithms (CCAs) has shifted from control-based to measurement-based in recent years.
However, we find that measurement-based CCAs, although having better performance, are not robust enough in fluctuating network environments, which are increasingly common nowadays.
In this paper, we propose PAD to make measurement-based CCAs as robust as control-based CCAs in fluctuating environments while enjoying the performance benefits in general.
PAD identifies that the root cause is that measurement-based CCAs blindly rely on measurement results, which unfortunately can be inaccurate, and will transiently mislead the CCAs to misbehave.
The preliminary design of PAD works as a shim layer between the socket and CCAs so as to scale to any measurement-based CCAs, which turns out to outperform most commonly used CCAs in fluctuating environments.


\end{abstract}

\maketitle

\section{Introduction}
The congestion control algorithm (CCA) community nowadays witnessed a significant shift in the design philosophy.
Traditionally, researchers and operators tended to conservatively increase or decrease the sending rate (or congestion window), which we call \textit{control-based} CCAs~\cite{cubic,vegas,copa}.
Such a design is helpful to stabilize the CCA -- the stability of such additive-increase-multiplicative-decrease (AIMD) algorithms has been analytically proved ~\cite{aimd}.
However, in recent years, increasingly more CCAs break the methodology of gradually increasing or decreasing the sending rate and take a \textit{measurement-based} way to make rate adaptation decisions~\cite{bbr,pcc}.
For example, BBR ~\cite{bbr} and PCC ~\cite{pcc} will deliberately probe the network bandwidth and directly take the measured value as the sending rate for the next step.
This is effective in the fluctuating Internet -- instead of slowly converging to the new bottleneck bandwidth in control-based CCAs, measurement-based CCAs can adjust the sending rate in one step.


However, one major issue of measurement-based CCAs is that they heavily rely on the measurement results, which unfortunately can be inaccurate and even misleading in some cases.
For example, the transient fluctuation of link propagation delay will aggregate the acknowledgment packets and affect the measurement results of available bandwidth.
The overestimation or underestimation of the available bandwidth will mislead the CCA and result in the overshoot or underutilization of the link (\S\ref{sec:motivation}).
Due to the mismatch between sending rate and available bandwidth, the CCA can not keep a stably high sending rate in the fluctuation situations, and degrades the performance.
Although not thoroughly investigated as us, similar observations about the robustness problem on specific CCAs have also been made by some previous researchers.
For example, when propagation delay is fluctuating heavily, BBR, a measurement-based CCA, may underestimate the round-trip propagation time ~\cite{ullbbr}.
In the above case, the network fluctuation makes it hard for measurement-based CCAs to estimate the network condition accurately. 
Moreover, we observe that such a performance degradation actually roots in the fundamental design flaws of the measurement-based CCAs in general.
These measurement-based CCAs probe the network with certain methods and collect some network samples to calculate the sending rate. 
In order to outline the network environment with just several instant samples, current measurement-based CCAs add some assumptions about the network environment.
Examples of such assumptions are that throughput can not exceed bottleneck bandwidth, and delivery rate can faithfully respect throughput. 
However, the assumptions may fail in dynamic situations like fluctuating delay situations, which means current measurement-based CCAs are not robust enough to face the complex network.

In light of the issue above, our question in this paper is: 

\textit{Can we have a robust measurement-based CCA while enjoying the performance benefits?}

We propose PAD as a trial of the robust measurement-based CCA. 
PAD comes from the following observation -- if the CCA can robustly measure the network environment, we can achieve both high performance and high flexibility. 
We add a stateful block for measurement-based CCAs to help them measure the network environment more robustly. 
Then, they can use both the historical information provided by PAD and the instant information provided by the measuring samples to generate the measuring result.

However, CCAs are heterogeneous and diverse, so it is challenging to make PAD general to all CCAs. 
We do not propose a new CCA directly, since there are a lot of measurement-based CCAs. 
An ideal solution is a plugin cooperating with all kinds of measurement-based CCAs. 
Re-arranging ACKs is a reasonable approach to send information to measurement-based CCAs since these CCAs usually use ACKs to generate measuring samples. 
Thus, PAD introduces an ACK controller between the TCP socket base (responsible for packet processing) and CCAs to help measurement-based CCAs get more robust measurement results. 
PAD is designed to keep historical information like ACK arrival timestamps, and inform the CCA of the information by re-arranging ACKs.
With the aid of the historical information provided by PAD, the CCA is expected to achieve better performance in dynamic situations.

Yet, it is non-trivial to re-arrange ACKs to a proper position and overcome all the following challenges. 

First, the rearrangement of ACKs in PAD might incur additional delays for specific packets.
Yet, PAD should not affect the network performance in stationary network conditions.
Therefore, on one hand, PAD should not add additional delay to many packets.
On the other hand, the delays added by PAD should not interfere the original working cycles of CCAs, otherwise the CCA will get confused about those unexpected delays.
It is challenging to only add limited delay to specific packets while making the CCA robust as discussed above.

In order not to insert extra delay, PAD is designed to have a positive mode and a negative mode, and functions only in the positive mode. Extra delay can be avoided by switching wisely between these two modes. 
PAD also uses the outgoing packets to avoid interfering the working cycles of the CCA. When the CCA is probing the network, PAD can detect it and allow the corresponding ACKs to go through more actively.

Second, PAD's rearrangement of ACKs should not be affected by CCAs' recovery process. 
Packet loss is inevitable in network and CCAs have their routine to recovery from the packet loss. 
It is challenging to make the rearrangement cooperate with existing recovery process well.

Packet loss can be divided into two categories. Some of them can be recognized by Fast Recovery. Introducing SACK ~\cite{sack} into PAD can successfully filter out the influence of this kind of packet loss. The other kind of packet loss is recognized after RTO. PAD deals with this kind of packet loss by cleaning its states and give up all its historical information. Then ACKs received before the packet loss will not influence those after the packet loss.

We implement PAD over BBR as a preliminary experiments.
Our method has got great results in our experiments of delay-fluctuating situations. 
PAD+BBR can work better than pure BBR, with about 1.6x throughput and about 0.5x extra delay. 
PAD+BBR has a better tradeoff between throughput and latency than BBR, Copa, Vegas, and Cubic. 
In the following sections, we also use BBR as an example to explain our insights. 



\section{Background and Motivation}

In this section, we first introduce the background of congestion control algorithms and scenarios with fluctuating propagation delay in \S\ref{sec:background}. We then motivate the design of PAD in \S\ref{sec:motivation} with our experiments on measurement-based CCAs in fluctuating delay situations, and our discussion about the experimental results.

\subsection{Background} \label{sec:background}

In this subsection, we first introduce control-based CCAs and measurement-based CCAs respectively. Then, we show that network scenes with fluctuating propagation delay are important nowadays.

\subsubsection{Congestion Control Algorithms}

We in this paper divide CCAs into two categories: control-based and measurement-based. They are divided by the means they acquire network information.
Control-based algorithms like Cubic ~\cite{cubic} and Copa ~\cite{copa} passively acquire network information. They have little knowledge of the network environment and passively accept congestion signals.
Measurement-based algorithms like BBR ~\cite{bbr} and PCC ~\cite{pcc} actively probe network information. 
They periodically enter a probing phase where they change their sending rate higher than the available bandwidth purposely to see how the network responds, and use the results to decide the sending rate. 

\textbf{Control-based CCAs:} Typically, control-based CCAs are believed to have worse performance and can work more generally ~\cite{expbbr}.
Control-based CCAs lack information about the network environment. 
They probe the network gradually based on their current states, and takes some action when congestion signals like packet loss or increasing latency emerge.
Such mechanisms mean that it is hard for a control-based CCA to achieve both high throughput and low latency. 
For example, Cubic is known to have high throughput and high latency, and Copa is known to have low throughput and low latency. 
However, since control-based CCAs do not rely on the perception of network environment,
they are robust to fluctuations in the network environment.

\textbf{Measurement-based CCAs:} Measurement-based CCAs perform better in traditional stable situations like wired network.
Measurement-based CCAs can actively probe the network environment. 
They can use collected information to decide the sending rate directly. 
If they can estimate the network environment well, they can decide the sending rate accordingly, and achieve both high throughput and low latency.
However, when measurement-based CCAs can not measure the network condition accurately in dynamic situations like fluctuating delay situations, they may fail to achieve their goals.
In other words, measurement-based CCAs are not robust enough to work in general situations.

\subsubsection{Propagation Delay Fluctuation}

Recently, there are a rising amount of network scenes with fluctuating propagation delay. 
Physical limits are the most important cause of delay fluctuation. Networks using mm-Wave as physical media like 5G is gradually going into people's sight ~\cite{ullbbr}. 
Utilization of low earth orbit (LEO) satellite is also actively applied ~\cite{leo}. 
In both situations, the delay fluctuation is rather high. For example, in LEO satellite network, the longest propagation delay can be twice than the shortest delay ~\cite{starperf}
Despite these physical limits, some techniques like delayed ACK and end-host or in-network scheduling overhead also lead to fluctuating delay. 

Since fluctuating delay situation is not a traditional stable network environment, we really want to know whether measurement-based CCAs can perform well in such a situation. In the next subsection, we will show that BBR can not perform well in fluctuation delay situations. We will deeply explain the reason and show our thinking about it.

\subsection{Motivation} \label{sec:motivation}

To figure out whether measurement-based CCAs can get good performance in fluctuating delay situations or not, we conduct a preliminary experiment using ns-3. We introduce some jitters to the link's propagation delay to see whether a CCA can work well on such links. We have the following observations.

\begin{figure}
    \centering
    \begin{subfigure}[b]{0.23\textwidth}
      \includegraphics[width=\linewidth]{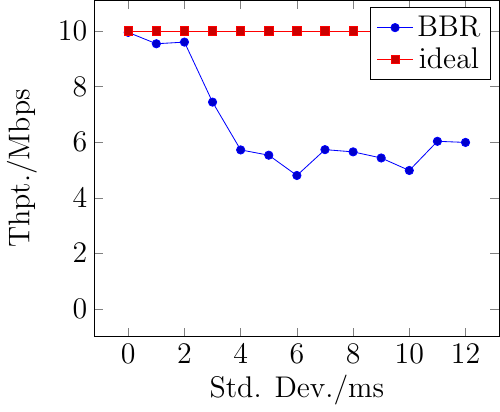}
      \caption{Throughput}
    \end{subfigure}
    \hfill
    \begin{subfigure}[b]{0.23\textwidth}
      \includegraphics[width=\linewidth]{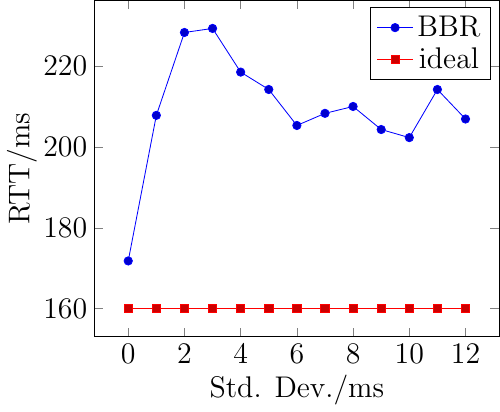}
      \caption{RTT}
    \end{subfigure}
    \caption{BBR does not perform well in fluctuating delay situations. With severer fluctuation, BBR can only achieve about half of the available bandwidth and introduce extra latency of more than 30 percent.}
    \label{fig:bbr}
\end{figure}

\textit{Measurement-based CCAs are not robust enough to face a fluctuating network environment and achieve both high throughput and low latency.}
For example, according to our preliminary experiments, a representative measurement-based CCA, BBR, does not work well in fluctuating delay situations. 
Details of the preliminary experiments are shown in \S\ref{sec:evaluations}.
As shown in Figure~\ref{fig:bbr}, BBR can take use of more than 99\% of available bandwidth while introducing extra latency of less than 7\% with a stable link. 
However, when the propagation delay starts to fluctuate, BBR can only use about half of the available bandwidth and introduce extra latency of more than 30 percent. That is to say, BBR can not work well in fluctuating delay situations.

\begin{figure}
    \centering
    \includegraphics[width=\linewidth]{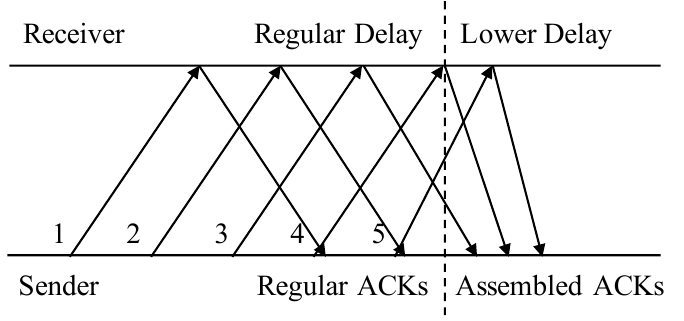}
    \caption{Demonstration for assembled ACKs. When the propagation delay becomes lower, ACKs will assemble.}
    \label{fig:accack}
\end{figure}


We will show the reason why BBR can not work well in fluctuation delay situations. We first show the mechanisms misleading BBR in \S\ref{sec:direct_cause}. We then show why measurement-based CCAs can not get accurate measuring results in \S\ref{sec:root_cause}. 

\subsubsection{Direct Cause} \label{sec:direct_cause}

When we dive into the detail of experiments, we find the fluctuating delay leads to BBR overestimating bandwidth, which then causes performance loss. The following several paragraphs will show the process in detail.

Fluctuating propagation delay causes assembled ACKs. 
Figure~\ref{fig:accack} is a demonstration of why fluctuating latency will cause assembled ACKs.
The sender sends all five packets with a constant pacing rate, and the first three ACKs come back with the same rate. When the fourth packet arrives the receiver, the propagation delay becomes lower. Then if we use ACK 3, 4, and 5 to calculate delivery rate, we will get a higher result.

Our preliminary experiments prove that BBR overestimates the bandwidth.
With the standard deviation to be 6ms, the 99 percentile and windowed-maximum percentile of all delivery rate samples are 8.5\% and 11.2\% higher than the available bandwidth respectively.
More detailed discussion can be seen in \S\ref{sec:evaluations}.


\subsubsection{Root Cause} \label{sec:root_cause}

In the above several paragraphs we showcase the reason why BBR does not work well in fluctuating delay situations. We can see the main problem is that BBR uses delivery rate samples to estimate bandwidth, but in fluctuating situations, the delivery rate samples do not faithfully reflect the bottleneck bandwidth. We believe this reveals a general drawback of all measurement-based CCAs.

The performance loss of BBR reflects the general drawback that measurement-based CCAs can not measure the network environment well in dynamic situations. 
Measurement-based CCAs probe the network in certain approaches and get some network samples. They then use these samples to calculate the proper sending rate. 
They are usually stateless, and therefore lack historical information.
It is a big challenge to use just instant samples to outline the network environment. 
Current measurement-based CCAs add some assumptions about the network environment to solve the problem. For example, BBR assumes throughput can not exceed bottleneck bandwidth, and delivery rate can faithfully respect throughput. Thus BBR uses the largest delivery rate (in several RTTs) as the estimation of bottleneck bandwidth. 
However, such kinds of assumptions can not always be true. In fluctuating delay situations, the delivery rate exceeds the bottleneck bandwidth from time to time, and thus BBR faces a performance loss in such situations.

We believe a better solution is to keep some states to store historical information. 
This idea is motivated by control-based algorithms, which use the inner state to decide sending rate when no congestion signal emerges. If we add some states to measurement-based CCAs, they can fully utilize the information they have collected during the probing phases. Then the newly designed CCA can use both the states and samples to generate the sending rate. The states keep track of outstanding historical information, and the samples reflect instant information. They can supplement each other, helping the CCA to generate a better sending rate.
\\ 


\textit{Question: Can we have a robust measurement-based CCA while enjoying the performance benefits?}
Existing methods suffer from the fundamental trade-off: control-based CCAs can provide robust performance, while measurement-based CCAs can provide high performance. In this paper, we are going to add a stateful block from control-based CCAs to measurement-based CCAs, enjoying the benefits from both sides.
With the aid of PAD, a CCA can be more robust while keeping its high performance. In \S\ref{sec:design}, we will introduce our PAD in detail. In \S\ref{sec:evaluations}, we will show some experimental results of PAD.

\section{Design} \label{sec:design}

In this section, we first give an outline of PAD in \S\ref{sec:design-overview}. 
We then introduce the design of PAD ACK Buffer and PAD Rate Estimator in \S\ref{sec:design-buffer} and \S\ref{sec:rate_estimator} respectively.

\subsection{Overview}
\label{sec:design-overview}

As we have discussed in \S\ref{sec:motivation}, measurement-based CCAs only use instant samples to measure the network environment. However, instant samples do not always faithfully reflect the network environment. 
PAD is then proposed to keep the historical information.

\begin{figure}
    \centering
    \includegraphics[width=\linewidth]{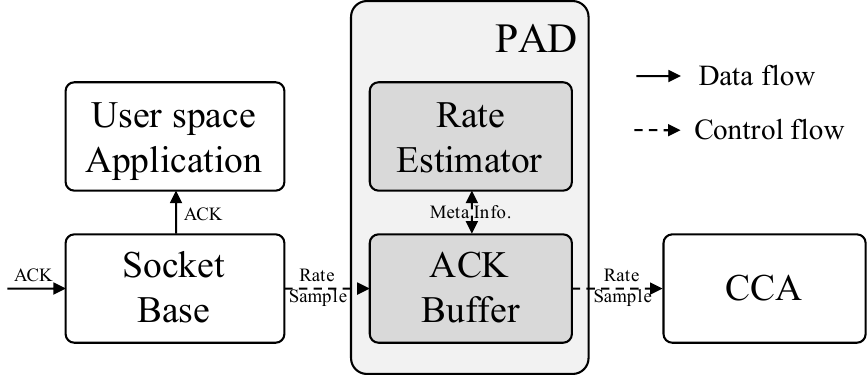}
    \caption{The structure of PAD, and its relationship with the TCP socket base and the CCA. PAD is composed of Rate Estimator and ACK Buffer. Rate Estimator stores the historical information, and ACK Buffer controls the ACKs directly. PAD works as a shim layer between the TCP socket base and the CCA.}
    \label{fig:PADstructure}
\end{figure}

PAD is an ACK controller, working between the TCP socket base and the CCA. 
The relationship among PAD, TCP socket base, and CCA is shown in Figure~\ref{fig:PADstructure}. Since the lack of historical information is the general drawback of many measurement-based CCAs, such modulated design can maintain as much flexibility as possible. Every new measurement-based CCA found to be influenced by the fluctuating delay situation can co-work with PAD, trying to get better performance.

PAD has two sub-modules, namely ACK Buffer and Rate Estimator. The structure of PAD is also shown in Figure~\ref{fig:PADstructure}.


ACK Buffer is where we insert historical information into measurement-based CCAs. As discussed above, PAD collects historical information for the CCA behind it. However, it is not an easy job to pass the historical information.
Since measurement-based CCAs often calculate the acknowledged bytes within a time window to form their probing samples, we decide to pass historical information by re-arranging the ACKs.
ACK Buffer can postpone received ACKs for some time to help the CCA get better samples.

Rate Estimator is where we collect and store historical information. Many measurement-based CCAs calculate ACK arrival rate to estimate the network environment, so we also use it as a concrete representation of the historical information. Rate Estimator monitors the ACK arrival rate, and instruct ACK Buffer to filter out some outliers.

\subsection{ACK Buffer}
\label{sec:design-buffer}

ACK Buffer is where arriving ACK waits for permission to go through PAD. 
When a rate sample represented for an ACK comes from Socket Base and is about to go into CCA, it will be queued in ACK Buffer first. ACK Buffer is a FIFO queue, with the job to keep ACKs temporarily. It reports to Rate Estimator when an ACK arrives and how many bytes the ACK acknowledges. Then it uses the ACK arrival rate got from Rate Estimator to decide when to allow the ACK waiting at the head of the queue to go into CCA.

Rate Estimator can calculate the ACK arrival rate $\lambda$, which will be shown in \S\ref{sec:rate_estimator}. ACK Buffer then uses calculated $\lambda$ to decide when to allow the next ACK to go through.
An easy way to do such a job is to choose another leaving rate $\mu$. $\mu$ represents the rate ACKs leave ACK Buffer. Since we want ACK Buffer to be empty most of the time rather than to keep a lot of ACKs, $\mu$ has to be chosen elaborately. In fact, $\mu$ should be slightly higher than $\lambda$ to keep ACK Buffer empty.
Specifically, ACK Buffer decides $\mu$ by
\begin{equation}
    \mu = k\lambda
\end{equation}
where $k$ is a parameter controlling the degree of aggressiveness.
When $k$ gets smaller, the ACK Buffer drains more slowly, which adds more extra delays.
When $k$ grows larger, the leaving rate are more different from the arrival rate, which might mislead the CCA behind the ACK Buffer.

With $\mu$ chosen, ACK Buffer can decide when to grant the next ACK to go through. 
The time to grant the next ACK to go through can be calculated by the pacing method with the pacing rate to be $\mu$.

\noindent\textbf{A critical goal of the design of PAD is not to inject additional latency in stable conditions. }
ACK Buffer introduces two modes to prevent injecting additional latency in stable conditions.
Not injecting additional latency means ACK Buffer should be empty in most of the time.
To achieve it, a positive mode and a passive mode are introduced to ACK Buffer. During passive mode, ACK Buffer allows every arriving ACK to go through ACK Buffer immediately. During passive mode, however, ACKs will be buffered for a period of time. 
In a stable network condition, ACK Buffer should work in passive mode, as if ACKs go from the socket base straight to the CCA. When the propagation delay starts to fluctuate, making some ACKs arrive earlier than supposed, ACK Buffer then changes to the positive mode and puts these ACKs back off to reasonable time.

\noindent\textbf{ACK Buffer should not interfere CCAs' regular probing process.}
Usually, a measurement-based CCA has a mechanism to measure the network environment periodically, so that it can react to network environment changes in time. One of the most popular approaches is to send more packets than the estimated bottleneck bandwidth on purpose. 
However, it is a challenge to distinguish the regular probing process from assembled ACKs caused by fluctuating delay, since there are both some ACKs crowding together.
In order not to interfere CCAs' regular probing process, ACK Buffer cooperates with the sending side of CCAs to distinguish CCAs' regular measuring process.
PAD can identify a period of time as the CCA's probing period when sending rate is higher than $\mu$, the rate CCA gets ACKs. 
If PAD and CCA can exchange some messages, then PAD can get the probing period directly from the CCA.
After PAD identifies a period of time as the CCA's probing period, PAD can 
allow more ACKs to pass when the corresponding ACKs come back. 
By such means, ACK Buffer will not block CCAs' regular measuring process, while pacing the arriving ACKs in the meantime.

\subsection{Rate Estimator} \label{sec:rate_estimator}

Rate Estimator is responsible for collecting and restoring the historical information. It also uses such historical information to estimate the ACK arrival rate, which will then be used to decide when to allow arrived ACK to go through. 
There are actually two assumptions here. First, we assume that the sender is trying to send as much data as possible. That is to say, the sender is not in the "application limited" state, where upper applications do not provide enough data to send. Second, the propagation delay of the link is fluctuating among a central value, and the central value is stable somehow.
If the above two assumptions can be true, the ACKs will arrive at a constant rate in general. Certainly, with the fluctuation, the arriving rate can not be exactly constant but is also fluctuating among a central rate. Rate Estimator's job is to find the central rate, which will be marked as $\lambda$ in the following paragraphs.

Rate Estimator continuously collects information to calculate proper $\lambda$. 
As we have discussed above, ACKs arrive at PAD at a constant rate in general. Thus, Rate Estimator uses a window to calculate the constant rate $\lambda$. More specifically, every time an ACK arrives, Rate Estimator upgrades $\lambda$ by
\begin{equation}
    \lambda = \frac{ACK_{now} - ACK_{prev}}{t_{now}-t_{prev}} \label{eq:calc_lambda}
\end{equation}
where $t_{now}$ is the time ACK arrives, $t_{prev}$ is $w$ RTTs before ACK comes, $ACK_{now}$ is the largest sequence number acknowledged by the newly arriving ACK, and $ACK_{prev}$ is the largest sequence number acknowledged by the ACK having arrived at $t_{prev}$. 
$w$ is a parameter controlling stable level. 
We set $w$ to 16 to make sure it can cover the probing period of commonly used measurement-based CCAs.

PAD uses the SACK option in TCP to resolve the ambiguity caused by ACKs.
When packet loss occurs, ACKs will introduce ambiguity. The acknowledged sequence number may not represent the largest sequence number received. 
SACK is a solution for such a problem. We read the SACK option field in the TCP packet header to get the largest sequence number received.
By getting the largest sequence number received more precisely, we can calculate $\lambda$ more precisely, and match the received ACKs to the packets sent more accurately.
Then we can decide the postponing time more precisely, and recognize the regular probing process more accurately.

\section{Preliminary Evaluation} \label{sec:evaluations}

\begin{figure}
    \centering
    \begin{subfigure}[b]{0.23\textwidth}
        \centering
        \includegraphics[width=\textwidth]{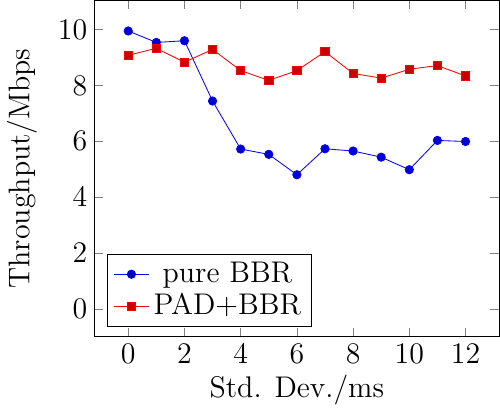}
        \caption{Throughput}
    \end{subfigure}
    \hfill
    \begin{subfigure}[b]{0.23\textwidth}
        \centering
        \includegraphics[width=\textwidth]{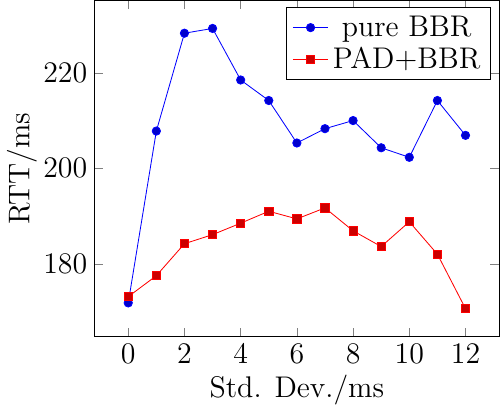}
        \caption{RTT}
    \end{subfigure}
    \caption{Comparisons between pure BBR and PAD+BBR. PAD+BBR can keep a stable throughput, while pure BBR faces a throughput collapse. PAD+BBR also introduces lower extra delay than pure BBR.}
    \label{fig:bbrvsPAD}
\end{figure}

\begin{figure*}
    \centering
    \begin{minipage} {0.63\textwidth}
    \begin{subfigure}[b]{0.49\textwidth}
        \centering
        \includegraphics[width=\textwidth]{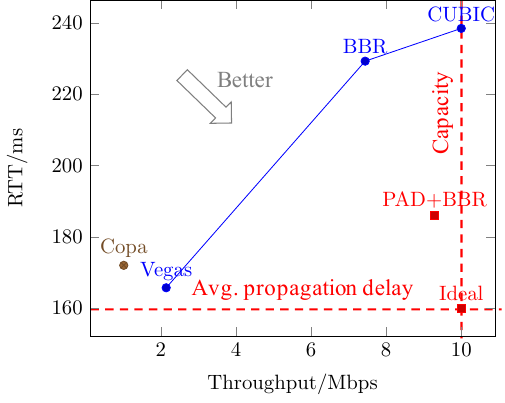}
        \caption{StdDev = 3}
    \end{subfigure}
    \begin{subfigure}[b]{0.49\textwidth}
        \centering
        \includegraphics[width=\textwidth]{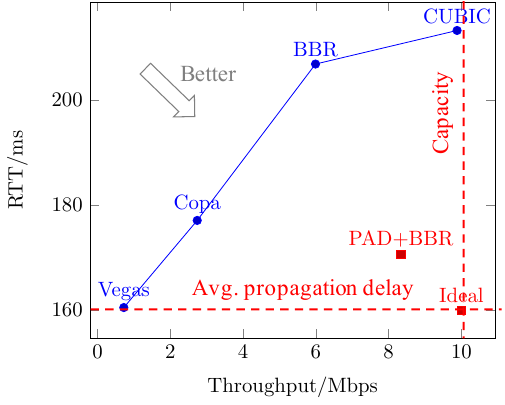}
        \caption{StdDev = 12}
    \end{subfigure}
    \caption{PAD+BBR surpasses the Pareto optimal line of throughput and RTT composed by pure BBR, Cubic, Vegas, and Copa in delay-fluctuating situations of different extents. The ideal throughput and RTT are also plotted as Ideal in the figures.}
    \label{fig:tradeoff}
    \end{minipage}
    \hfill
    \begin{minipage} {0.32\textwidth}
        \includegraphics[width=\textwidth]{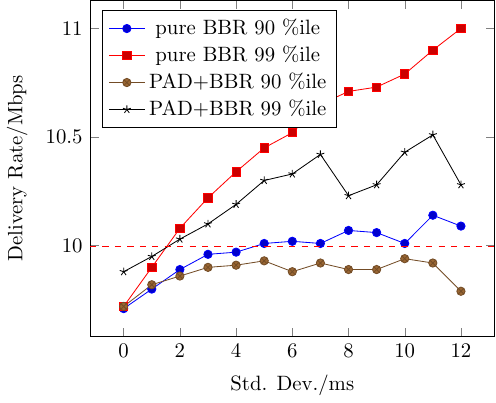}
        \caption{Percentiles of the delivery rate samples. PAD can alleviate the extent that BBR gets samples of delivery rate higher than available bandwidth.}
        \label{fig:percentiles}
    \end{minipage}
\end{figure*}

Since PAD works between the TCP socket base and the CCA, it is hard to evaluate PAD itself. As we have mentioned in \S\ref{sec:design-overview}, if we see PAD and CCA as a whole, it can be seen as a new CCA. Thus, we use the combination of PAD+BBR as a new CCA and compare it with other CCAs.

We first compare PAD+BBR and pure BBR to verify the improvement PAD can achieve. Then we compare PAD+BBR with BBR, Cubic, Copa, and Vegas. We compare both throughput and RTT among these CCAs. After that, we dive into BBR to get some inner information to prove that our explanation of BBR's performance loss is reasonable. 
In the end, we show PAD can provide improvement in more complicated situations, and will not introduce fairness problems.

Most of our experiments discussed below are conducted with ns-3~\cite{ns3}.
We use ns-3 to simulate a sender, a bottleneck link, a delay-fluctuating link, and a receiver. 
The delay-fluctuating link is simulated by changing the link propagation delay every 100 milliseconds during the whole experiment. 
The propagation delay's changing process is specially smoothed to keep the order of packet arrivals. 
In general, the propagation delay is normally distributed, and we control the standard deviation to get different extents of fluctuation.
The bandwidth of the bottleneck link is set to 10Mbps, and the round-trip propagation time is set to fluctuate around 160ms. The queue at the bottleneck is a pure FIFO queue with a length of 100 packets. For PAD, we set $k$ to 1.025, and $w$ to 16.

PAD+BBR performs better than pure BBR in terms of both throughput and latency.
Since we can see PAD+BBR as an improvement of pure BBR, the first experiment we conduct is to compare them.
The two sub-figures of Figure~\ref{fig:bbrvsPAD} show the throughput and RTT of PAD+BBR and BBR in different extents of delay fluctuation respectively.
When the extent of the fluctuation grows severer, BBR can only utilize about half of the bottleneck bandwidth, with extra latency several times than the theoretical value. With the aid of PAD, PAD+BBR can keep a stably high throughput and lower RTT.
This result in a sense proves our explanation about why BBR dose not perform well in delay-fluctuating situations. PAD successfully alleviates the phenomenon of assembled ACKs and bandwidth overestimation by introducing historical information. 

PAD+BBR has got a better trade off than Cubic, pure BBR, Copa, and Vegas in different extents of fluctuation.
When evaluating a new CCA, throughput and RTT are the two most significant target. We supply the sender with all kinds of CCAs, and monitor the average throughput and RTT during the process. We choose Cubic, Vegas and Copa as representatives of control-based CCAs. Cubic uses packet loss as congestion signal, and Vegas and Copa use delay increase as congestion signal. We choose pure BBR (BBR without PAD) as a representative of measurement-based CCAs.
Our results are shown in Figure~\ref{fig:tradeoff}. It shows different CCAs' throughput and RTT in different fluctuation extent, namely with standard deviation to be 3ms and 12ms. The blue line across the image from top right to bottom left shows the Pareto optimal of throughput and RTT, with every point on it represents a CCA. Since we expect higher throughput and lower RTT, points at southeast are the best. It is easy to see PAD+BBR surpasses the Pareto optimal. It achieves rather low RTT with only a bit throughput loss.


The following experiments dive into how BBR is mislead by the rate samples.
We conduct the next experiments to show that BBR actually gets some improper delivery rate samples, and PAD alleviates the phenomenon by re-scattering assembled ACKs.
We present the distribution of the 90th and 99th percentiles of delivery rate estimations in BBR.
To show the influence of BBR's windowed maximum filtering approach, we also calculate the windowed maximum 90th and 99th percentiles.
The results are shown in Figure~\ref{fig:percentiles}. With fluctuation growing severer, percentiles of BBR, especially windowed maximum percentiles grow a lot. PAD, however, manages to keep all these percentiles or windowed maximum percentiles lower. Thus, PAD avoids BBR from samples with delivery rate of too high. This can avoid BBR from overestimate the bottleneck bandwidth.


PAD+BBR also performs better than pure BBR when at least two streams exist.
It is a more common situation that at least two streams flow through the same bottleneck link. 
We add different numbers of flows to the bottleneck link. Specifically, experiments are conducted on 2 and 5 streams. The flows are added one by one, with 0.1 seconds between two flows. 
The experiments show that PAD+BBR can always get higher throughput than pure BBR in different extents of fluctuation. The increase of throughput varies from 1.06x to 1.46x.


The fairness between PAD+BBR and other CCAs are also achieved.
It is important to confirm the newly proposed method will not harm existing methods. Thus, we put two streams into the bottleneck link, both using BBR as the congestion control algorithm. One of the two streams is armed with PAD, while the other one is not. 
Results show that PAD+BBR can coexist with pure BBR, without any one of them facing the danger of starvation. 
The stream with higher throughput takes up less than 1.11x throughput than the other stream.
That is to say, PAD+BBR can easily be deployed, even when many pure BBR streams are still in the Internet.

\section{Discussion}

We present several future directions beyond the preliminary design of PAD.

\textit{Will PAD work with other CCAs?}
In this paper, we only evaluate the performance of PAD over BBR.
However, PAD is designed to work with any other measurement-based CCAs.
As long as the CCA depends on measurement results (e.g., delivery rate or latency), PAD can help to make the measurement results robust by introducing the queue.
For example, PCC probes the network periodically to get an instant sample of delivery rate and loss rate. 
It is possible that PAD may improve PCC's performance since the mechanism is almost identical to BBR except for the algorithm.
In the future, we will implement and evaluate PAD over other CCAs.

\textit{Is PAD easy to use for network operators?}
Since we are modifying part of the network stack of Linux kernel, a natural concern is if it is too difficult for network operators to use PAD in their own products.
In fact, PAD takes the load of modification from the users of PAD -- network operators will not need to touch either the CCA or the kernel codes themselves.
For example, we can insert a kernel module to deploy PAD.
As long as operators can insert the PAD module into their own operating system, PAD should work as expected.
We plan to implement this into a kernel module for a broader impact in the future.

\textit{The influence of measurement-based CCAs on existing modelling.}
Since the proposal of BBR~\cite{bbr}, we do see an inspiring spike in the CCA research by heavily relying on measurement results, most of which do have a satisfactory performance.
However, measurement-based CCAs are not well analytically studied in the community: the robustness we discussed in this paper is one aspect, but definitely not the only one.
For example, measurement-based CCAs such as BBR will also break the throughput model of existing control-based CCAs~\cite{bbrmodel}.
There are definitely more exciting directions to explore, especially considering the fact that the measurement-based CCAs are gradually dominating the Internet traffic~\cite{tcpcc,bbrdominant}.
We call for the attention from the community to rethink these designs and their potential effects together.

\section{Conclusion}

In this paper, we propose PAD to collect historical information for measurement-based CCAs. 
PAD works between the socket base and the CCA. It collects and stores the historical information, and then passes to the CCA by re-arranging the ACKs.
We conduct some preliminary experiments to show PAD can work well with BBR, one of the most representative measurement-based CCAs.


\bibliographystyle{ACM-Reference-Format}
\bibliography{sample-base}

\end{document}